\documentclass[sigconf]{acmart}

\usepackage{multirow}
\usepackage{array}
\usepackage{graphicx}
\usepackage{CJKutf8}
\AtBeginDocument{%
  }

\setcopyright{acmlicensed}
\copyrightyear{2024}
\acmYear{2024}
\acmDOI{XXXXXXX.XXXXXXX}

\acmConference[Chinese CHI 2024]{The Twelfth International Symposium of Chinese CHI}{November 22~25, 2024}{Southern University of Science and Technology, Shenzhen, China}

\newcommand{\eg}{{\it e.g.,\ }}

\newcommand{\name}{{\it VChatter}}
\newcommand{\pzh}[1]{{\color{black} #1}}
\newcommand{\peng}[1]{{\color{black} #1}}

\newcommand{\zh}[1]{{\color{black} #1}}

\begin{document}
\begin{CJK}{UTF8}{gbsn}
\title{VChatter: Exploring Generative Conversational Agents for Simulating Exposure Therapy to Reduce Social Anxiety}

\author{Han Zhang}
\email{zhangh773@mail2.sysu.edu.cn}
\affiliation{%
  \institution{Sun Yat-sen University}
  \city{Zhuhai}
  \country{China}
}

\author{KaWing Tsang}
\email{kawing-sd.tsang@connect.polyu.hk}
\affiliation{%
  \institution{The Hong Kong Polytechnic University}
  \city{Hong Kong}
  \country{China}
}

\author{Zhenhui Peng}
\email{pengzhh29@mail.sysu.edu.cn}
\affiliation{%
  \institution{Sun Yat-sen University}
  \city{Zhuhai}
  \country{China}
}

\renewcommand{\shortauthors}{Zhang et al.}


\begin{abstract}

\pzh{
Many people struggle with social anxiety, feeling fear, or even physically uncomfortable in social situations like talking to strangers. 
Exposure therapy, a clinical method that gradually and repeatedly exposes individuals to the source of their fear and helps them build coping mechanisms, can reduce social anxiety but traditionally requires human therapists' guidance and constructions of situations. 
In this paper, we developed a multi-agent system VChatter to explore large language models(LLMs)-based conversational agents for simulating exposure therapy with users. 
Based on a survey study (N=36) and an expert interview, VChatter includes an Agent-P, which acts as a psychotherapist to design the exposure therapy plans for users, and two Agent-Hs, which can take on different interactive roles in low, medium, and high exposure scenarios.
A six-day qualitative study (N=10) showcases VChatter's usefulness in reducing users' social anxiety, feelings of isolation, and avoidance of social interactions.
We demonstrated the feasibility of using LLMs-based conversational agents to simulate exposure therapy for addressing social anxiety and discussed future concerns for designing agents tailored to social anxiety.
}

\end{abstract}




\maketitle
\section{Introduction}

\pzh{
Adolescents, being a more psychologically sensitive group, are more susceptible to mental health disorders such as depression and anxiety \cite{kendall2015}. Xin et al. \cite{xin2022} reveal that due to societal changes, an increasing number of adolescents in China are suffering from Social Anxiety Disorder(SAD). They feel fear or even physically uncomfortable (\eg excessive blushing and palpitations \cite{morrison2013}) in social situations like talking to strangers and attending to a party.
These experiences force individuals to withdraw from social interactions \cite{stein2001} or adopt safety behaviors (such as focusing only on themselves) to protect themselves, leading to a cessation of communication with others \cite{heimberg1995}. 
However, relationships and communications between individuals are fundamental needs \cite{taormina2013}, which is crucial for our well-being \cite{ryan2000}. Yet social isolation caused by social anxiety often exacerbates the feelings of disconnection and isolation \cite{cacioppo2009}, potentially leading to further mental health issues, such as depression \cite{loades2020}. 
To treat the disorder, exposure therapy is widely used in the clinic, which has a good effect on social anxiety and other anxiety disorders, including OCD and PTSD~\cite{heimberg1985, ponniah2008}. 
The core concept of exposure therapy is to place the 
patients in real situations or stimulate occasions that cause anxiety or fear, so that patients can gradually get used to and accept their fear, thereby reducing avoidance behaviors and lowering the intensity of anxiety or fear \cite{field2015}. 
Exposure therapy for those with social anxiety typically includes guided social interactions, such as participating in conversations or attending social events (e.g., job interviews, and meetings) to help them learn, adapt, and confront their uncontrolled fear \cite{heimberg2002}. However, engaging patients in real-life social scenarios can be challenging, as individuals often distance themselves mentally during treatment, which reduces their engagement and effectiveness of exposure therapy treatment \cite{hope2010}.
In real-life exposure therapy, some patients may also experience fear or resist directly engaging in real-life situational interactions.
Additionally, the stigma associated with seeking professional therapy may prevent individuals from accessing treatment \cite{sad2000, national2013}. 

To address these challenges, previous Human-Computer Interaction (HCI) researchers have explored providing users with a safer therapeutic setting in a virtual environment, typically through virtual reality (VR) devices ~\cite{krishnappa2022, zamanifard2023, zhao2020}.
For example, Mello et al. \cite{krishnappa2022} utilized VR devices to create Virtual Human Agents (VHAs) to interact with users and then study the interplay of gaze and interpersonal distance to reduce social anxiety disorder. 
Although VR-based virtual environments can provide users with immersive experiences, VR devices are expensive and not accessible to many users. 
}

\pzh{
One promising alternative is using text-based or voice-based conversational agents (CAs) to simulate exposure therapy with users to reduce their social anxiety. 
Users can access the CAs via multiple devices like mobile phones and personal computers. 
Existing HCI works have proposed various CAs that help users handle mental health concerns ~\cite{weizenbaum1966, ng2023, lee2017}. 
For example, Schroeder et al. introduced  Pocket Skills \cite{schroeder2018}, a mobile app that teaches users Dialectical Behavior Therapy (DBT) via a CA to decrease users' depression and anxiety. 
However, these previous CAs largely rely on hand-coded dialogue flows and materials to play their roles in mental health interventions. 
When it comes to exposure therapy, previous CAs may fall short in constructing different social scenarios that users feel anxious about and playing the roles of social actors in those scenarios. 
Recent advances in large language models (LLMs) show great potential to enable CAs to simulate human roles in social contexts ~\cite{park2023,weizenbaum1966, ng2023, lee2017}. 
Nevertheless, little is known about the design, feasibility, and user experience of LLM-powered CAs for simulating exposure therapy to facilitate users' social anxiety. 
}

\pzh{
In this paper, we designed, developed, and evaluated a system named \name{} to explore generative conversational agents for simulating exposure therapy with users with social anxiety. 
We first distributed a questionnaire to 36 individuals who self-reported social anxiety to investigate the challenges when confronting their fears and to understand their perceptions and requirements for virtual agents. 
\peng{The results show that agents should be outgoing, gentle, and willing to listen. We also conduct an interview with an expert to discuss how to design a conversational agent for social anxiety individuals, and raised 4 design requirements.}
Based on these findings and suggestions, we designed and implemented \name{}, which consists of a psychotherapist agent \textit{Agent-P} and multiple social agents \textit{Agent-H} customized based on the social contexts in which users feel anxious. 
\peng{In this process, Agent-P, which uses LLM ChatGPT, will identify the source of each user's fear and develop a personalized exposure therapy plan, along with providing recommendations. Agent-H will then assume various interactive roles based on the exposure scenarios designed by Agent-P with the help of LLM. Additionally, during interactions, all agents' textual outputs will be converted to voice output in real-time by a text-to-voice model.}
We evaluated VChatter's effectiveness and user experience via a six-day study with 10 participants. 
\peng{
The results indicate that users experienced a significant reduction in social anxiety after undergoing our virtual exposure therapy, along with a noticeable alleviation of loneliness. VChatter, by following the steps of exposure therapy, helped users build positive behavioral and cognitive patterns, facilitating their return to normal social interactions. The use of virtual humans in the exposure plans not only enhanced user immersion but also provided a safer exposure environment. Based on user feedback, we offer two suggestions for future researchers.
}
}

\begin{figure*}[!htb]
  \centering
  \includegraphics[width=\textwidth]{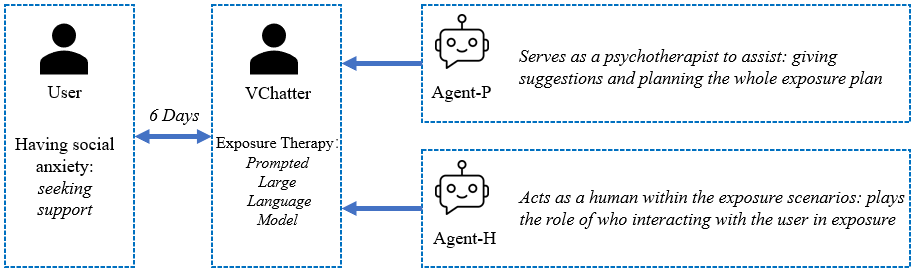} 
  \caption{Design of VChatter}
  \label{fig: flow}
\end{figure*}

\section{Related Work}

\subsection{Social Anxiety Disorder}
Social Anxiety Disorder is an anxiety disorder marked by a persistent fear of social or performance situations when patients in social occasions \cite{sad2000,kampmann2016}. Patients often experience involuntary trembling, blushing, and anxiety during social interactions, which leads them to avoid or withdraw from social activities \cite{stein2001}. Social relationships are essential, and  a prolonged lack of social interaction can result in social isolation, which may then lead to other mental health issues, such as depression \cite{teo2013}. Zhang et al. \cite{loades2020} investigated the mental health of adolescents in isolated environments during COVID-19 and found that nearly half experienced intense loneliness, which shows loneliness was strongly correlated with their mental health. For young individuals, SAD can also lead a freezing or mutism \cite{sad2000}.
In clinical, Cognitive Behavioral Therapy (CBT) in exposure therapy is the primary treatment for Social Anxiety Disorder \cite{field2015}, which is more effective and avoids the risks of medication dependency \cite{sareen2000}.
Exposure therapy is the commonly used behavioral treatment, where individuals repeatedly confront their fear triggers to learn how to cope and process their fears more healthily and safely \cite{heimberg1985}. One of the widely used concepts of the treatment process is CBT. The concept of CBT is helping individuals with social anxiety by developing positive mental awareness and linking it to appropriate behaviors, addressing underlying psychological issues. Depending on the specific case of social anxiety, CBT can help patients gradually gain control over their fear of social situations and attempt to transform that fear into other emotions, such as confidence.
Unfortunately, many patients hesitate to seek treatment due to the 
stigma \cite{national2013}, feelings of shame \cite{sad2000}, or overwhelming fear form themselves \cite{stein2001}. For those who do participate in exposure therapy, fear of being judged \cite{Lee2020} or showing vulnerability to others \cite{Lee20cscw} can also affect treatment effectiveness. This reluctance to fully participate in the therapeutic process may hinder their ability to confront their fears and ultimately limit their progress.


\subsection{Conversational agents Designed for Mental Health}

In the field of Human-Computer Interaction (HCI), many researchers explored using a more convenient and safer way to help individuals with mental health concerns alleviate or treat their issues. A common approach is to develop conversational agents (CAs) that provide support. The first chatbot, ELIZA \cite{weizenbaum1966}, was designed to emulate a psychotherapist by offering empathetic responses to patients with psychological disorders. However, as a code-based chatbot, the logic of the program significantly limits its ability to interact with users.
Peng et al. \cite{peng2020} developed MepsBot to assist users in online mental health communities with self-disclosure. This retrieval-based agent is user-friendly and responds quickly, but it struggles to handle the complex logic often present in conversations as it is a retrieval-based chatbot. Besides, Li et al. \cite{Li24} Using conversational agents to assist the communication between patients and therapists, highlighting the convenience of these agents and underscoring the benefits of a blended care approach.
With the advancements in natural language processing, CAs have shown significant improved logical capabilities and  potential in mental health therapy. For instance, Ng et al. \cite{ng2023} created a culturally adaptive chatbot based on the BERT model, which effectively reduced the stress of users with depressive tendencies. Lee et al. \cite{Lee2023} employed a conversational agent to simulate a psychiatric patient, fostering a social connection with users and helping to reduce the stigma associated with mental illness. In another study, Lee et al. \cite{lee2017} developed a CA using Manychat and Google Dialogflow, demonstrating that building a good relationship with users encourages them to share more about themselves, ultimately helping to relieve stress and promote positive interpersonal relationships.
The performance of chatbots has improved alongside advancements in logical modeling and is positively correlated with user experience. Currently, large language models (LLMs) like GPT exhibit strong logical capabilities in daily conversations. However, there remains a gap in utilizing these LLMs to address specific mental health issues. Therefore, we aim to design a conversational agent using the mainstream LLM ChatGPT to assist individuals with social anxiety.


\subsection{Virtual environment for therapy}

A common approach in mental health therapy is using virtual reality (VR) to create a safe and controlled virtual treatment environment for users. For example, Heng et al. \cite{heng2021} developed ReWIND, a role-playing game designed to help anxiety patients manage their emotions in a virtual setting, thereby reducing their worries and fears. On other hand, Pradeep et al. \cite{krishnappa2022} created a VR-based communication platform that analyzes the facial expressions of individuals with autism spectrum disorder (ASD) or social anxiety disorder (SAD) to understand their emotions better. Zamanifard et al. \cite{zamanifard2023} also built a virtual social environment using VR to assist users in overcoming their fear of social interactions.
Previous studies suggest that a virtual social environment can provide a safer environment for anxiety patients to engage in therapy while also offering a higher level of immersion. However, compared to conversational agents, VR-based tools often require greater financial investment and human resources.
Additionally, promoting these tools may be challenging if users are unfamiliar with the equipment. In contrast, CAs have demonstrated effective results in mental health therapy, providing advantages such as reduced costs and fewer obstacles to access.
Considering that interactive virtual factors, such as character and scene, significantly enhance the immersion \cite{mal2024}, we propose integrating virtual human agents and virtual scenes into our conversational agents to create an improved immersive experience.
\zh{
The advantages of VR in creating a more realistic virtual environment are clear: it provides higher immersion and more authentic interactivity. However, there are also notable drawbacks. VR often requires users to interact face-to-face with researchers, and not every participant may own VR equipment. Additionally, since our research focuses on specific individuals, we need to consider whether they might engage in avoidance behaviors. To create a safer and more accessible virtual environment, we explored how to build virtual scenes within VR devices using a chatbot interface.
}

\section{Survey Study}

To understand the current situation of social anxiety among adolescents and non-clinical methods for their self-coping,
we distributed a survey on a local campus network forum to gather teenagers' attitudes and suggestions regarding using chatbots to assist in treating Social Anxiety Disorder. The survey first required participants to complete the Liebowitz Social Anxiety Scale (LSAS) and report their Liebowitz Social Anxiety Score. Then they were categorized into two groups based on their scores: potential SAD individuals ($30 \leq LSAS < 60$), and clinically diagnosable SAD individuals ($LSAS \geq 60$). We received 36 responses in total, 19 had reported an LSAS of at least 60, who were clinically diagnosable as having social anxiety disorder.

For participants with a LSAS score of 60 or higher (n=19), we first explored the common scenarios that typically trigger their anxiety. We then inquired about how this anxiety has affected their daily lives. The responses revealed that six participants felt anxious in public places, seven experienced fear during presentations, three reported difficulties interacting with strangers, and three mentioned other situations, such as responding to jokes.
Next, we asked them what types of support they believed could help alleviate their anxiety. Two participants mentioned the help of medication, nine expressed a desire to overcome their anxiety through trying repeatedly, three hoped for encouragement from others (e.g. families and friends), six preferred to avoid social situations, and six reported having no idea how to improve their condition.

We used a 7-point Likert scale to evaluate participants' perceptions of whether a virtual agent capable of engaging in conversation could be beneficial for social anxiety suffers. The average score for individuals who may have SAD was 6.33, while the mean score for those with clinically diagnosable SAD was 4.74. We then asked participants who supported the idea of a virtual agent to identify the characteristics they would like the agent to possess. Out of 36 responses, 30 participants indicated preferences, with a score of 3 or higher on the scale. According to their feedback, an effective virtual agent for assisting individuals with SAD should primarily exhibit the following qualities: being outgoing and gentle (14 out of 30), a willingness to listen (7 out of 30), and patience (6 out of 30). These characteristics will inform the design of our agent prompts.
Furthermore, participants who felt that this approach was ineffective showed a reluctance to engage in any form of social activity. Among all participants, 6 out of 36 rated their inclination to use the virtual agent at less than 3. Their LSAS scores were reported to be over 60, and they also rated their interest in this scale as 1 point. At the end of survey, they clearly expressed their refusal to participate in any form of social interaction.
Overall, the responses indicated that these scenarios primarily involve public settings, such as giving presentations, making new friends, and attending social gatherings.


To better design our agent and experiment flow, we conducted a one-hour discussion with a certified psychology expert(male, age 28) who holds a mental health first aid certificate issued by MHFA \footnote{MHFA is an evidence-based, early-intervention course that teaches participants about mental health and substance use challenges (\url{https://www.mentalhealthfirstaid.org/}).}. 
We first discussed which method could be implemented based on chatbots to assist individuals with social anxiety.
\zh{
In our brainstorming, we also considered integrating chatbots into VR environments. However, we recognized that requiring users to use VR headsets in face-to-face settings might introduce additional stressors (e.g., gaze \cite{krishnappa2022}) that could lead to potential biases and increase avoidance behaviors. As a result, we opted for a more accessible and less intimidating online chatbot format. To maintain immersion, we retained key elements from VR, such as virtual avatars \cite{mal2024} and voice output \cite{Eagle2022}, which enhance user immersion.
Regarding anxiety reduction, we reached a consensus to leverage the well-established Cognitive Behavioral Therapy framework to guide our system's design. After ruling out the overly aggressive approach of flooding therapy, we chose a more gradual, controlled exposure therapy approach. This method allows users to become desensitized to anxiety triggers in a safer, step-by-step manner, ensuring a more supportive experience.
}
After brainstorming, we decided to model the chatbot interaction on the mainstream therapeutic approach for social anxiety disorder: exposure therapy. Exposure therapy is currently considered one of the most effective and safer approaches for SAD treatments. The therapy involves encouraging patients to engage with others in various scenarios to desensitize them to their fears while helping them rebuild positive awareness with the support of a therapist.
Then, in designing the chatbots with exposure therapy, the expert mentioned that previous work demonstrated the efficacy of chatbots in playing the role of therapist, the chatbot can be designed to guide participants through the therapeutic process as a therapist. Chatbots should also have the capability to play other roles encountered in real exposure scenarios to interact with patients. Moreover, in real-life exposure therapy, users engage in actual interactions. However, a text-based chatbot may not provide sufficient immersion for users.


For the design of agents' output, the expert also provided several key Design Requirements (DRs) we need to address in our design:
Firstly, as an agent designed to mimic a psychotherapist, it needs to guide users, which is a critical role of therapists in actual therapy. This includes, but is not limited to, helping patients identify the sources of their fears and guiding users through their treatment plans. Secondly, if the system includes multiple agents, their interactions should be more cohesive, for instance, taking each other's outputs into account; otherwise, this could create a disjointed experience for the user. Lastly, the user experience should follow the genuine steps of exposure therapy to achieve the desired outcomes of CBT. Based on expert suggestions and user survey results, we have summarized several design requirements as following:
\begin{itemize}
    \item \textbf{DR 1:} The agents must engage in conversations that help uncover the participant's fears and assist in identifying their issues, similar to how a human therapist would assess a participant's condition —- following the suggestions of the expert.
    \item \textbf{DR 2:} The agents will play both the therapist's role and other characters in real exposure scenarios \cite{heimberg1985}. and the connection between those agents should be tight. For the agent acting as the therapist, it needs to assist users in identifying the sources of their fears and guide them through the entire treatment process, which can fit the expert suggestions. For agents playing other roles, their personalities should be friendly and outgoing to encourage greater user participation in the chat, which can fit the users' need. To ensure a cohesive connection among agents, they should consider incorporating parts of each other's outputs as their inputs.
    \item \textbf{DR 3:} It is essential to establish clear stratification within the exposure hierarchy, and the design of exposure scenarios may need to be preconfigured \cite{Legaspi2023} or referenced against questions of the LSAS \cite{mennin2002}.
    \item \textbf{DR 4:} The agent should offer more than just text output to provide users with a more realistic experience and immersion \cite{Eagle2022, Cooney2024}.

\end{itemize}

\section{Design and Implementation of Vchater}
\subsection{Principle}
Based on the results of the user survey and suggestions of the expert, agents need to function as therapists by providing guidance and suggestions to users, while also acting as conversation partners that embody traits preferred by the users, which is crucial for encouraging user interaction. We categorized the agents into two types, which can facilitate \textbf{DR 1}:
\begin{itemize}
    \item Agent-P: Serves as a psychotherapist to assist the user throughout the entire experiment. There is only one Agent-P for every user.
    \item Agent-H: Acts as the interactive human within the exposure scenarios. The number and function of Agent-H vary depending on the specific exposure scenario.
\end{itemize}

As we see LLM-based agents have a good performance in guiding users in mental therapy \cite{mal2024}. We designed Agent-P, who acts as a human psychotherapist to guide the whole exposure therapy. To address \textbf{DR 3}, we designed the entire interaction process of Agent-P to follow the authentic exposure therapy protocol:

\textbf{Step 1 exposure therapy: Assessment.} Agent-P initiates the experiment by assessing the level of users' social anxiety that followed the LSAS. Agent-P then guides the user in exploring the source of their anxiety. Based on the identified fear source and anxiety level, a customized exposure plan is devised. Noticeably, each user's exposure plan will be structured over six days, during which users are required to sequentially experience low, medium, and high exposure scenarios twice.
\textbf{Step 2 Hierarchy Construction} The user will be asked by Agent-P to interact with Agent-H in scenarios specifically tailored by Agent-P and complete the task assigned by Agent-P. Users will be required to complete exposure tasks within the different exposure scenarios designed by Agent-P, with the difficulty of the tasks being positively correlated with the level of exposure scenarios (for example, in a mild exposure scenario, users may be asked to order food from a server of the opposite sex in a café).
\textbf{Steps 3 Expose step by step, Repeat, and Enhance.} After completing each goal, the user interacts with Agent-P again to discuss the challenges and difficulties encountered during the exposure. Based on the user’s progress, the next step of the exposure plan is refined and reinforced, continuing until the entire exposure plan is completed.
\textbf{Step 4 Skill teaching.}It is noticeable that if the user encounters difficulties during the exposure scenarios, they can seek assistance from the agent at any time. When designing exposure scenarios and setting exposure goals for the user, the agent also provides helpful hints to aid the user in completing the tasks.
\textbf{Step 5 Summary.} When all exposure tasks are completed, a summary of the user's performance and recommendations for future actions are provided.

In the exposure scenarios, users are required to interact with each of them individually or engage with both simultaneously in group tasks. The foundational design of both agents, Agent-P and Agent-H, adheres to the results of the user survey, meaning they are skilled listeners, approachable, and patient, with the ability to engage in lengthy dialogues. Additionally, Agent-H is designed to incorporate flexible elements: characteristic and interaction scenarios. Building on this foundational design, Agent-H will adjust its role based on the input character settings (e.g., as a peer or supervisor) and alter its responses according to the specified interaction scenarios (e.g., classroom or café). It is noteworthy that the character settings and interaction scenarios are customized by Agent-P based on user responses and the current stage of therapy. Furthermore, due to the memory limitations of the large model, we do not consider transferring memory from agent-h to agent-p; instead, agent-p will guide the user to summarize each interaction with agent-p, followed by providing feedback and suggestions to facilitate the next steps in therapy. These designs align with the specified \textbf{DR 2}.

About \textbf{DR 4}, we observed the immersion virtual human agents bringing \cite{mal2024} and the convenience voice agents provide \cite{Cooney2024}. We first considered preparing a virtual human representation for each of our agents, allowing this representation to have speech output, which means transforming the text output into spoken language. In other words, both Agent-P and Agent-H will read the content aloud to users while providing text output. Furthermore, the facial expressions of the virtual representation and the tone of voice during the reading should also vary following the content of the text. specially, Agent-H has been designed with two types, male and female, which have corresponding output and avatars. As for the style of virtual humans, to avoid infringement of image rights and the uncanny valley effect \cite{Mori12}, we opted for open-source animated models.


\subsection{Implementation}
\begin{figure*}[!htb]
  \centering
  \includegraphics[width=1\textwidth]{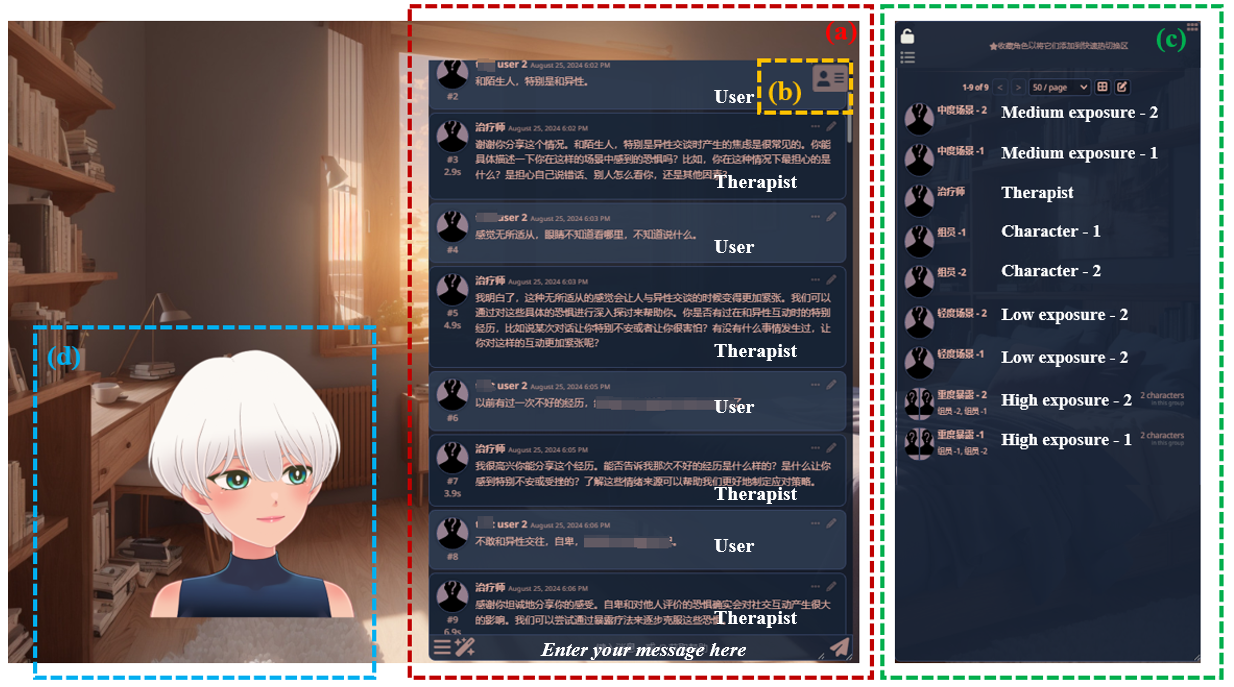} 
  \caption{(a) Chat history between the user and the therapist, with dark colors representing the therapist and light colors for the user.\textit{Red box.} (b) A history of all scenarios the user has interacted with, \textit{area (c)}, which will appear on the right side of the webpage when this button was clicked or will be hide when button was clicked again.\textit{Yellow box.} (c) Six exposure scenarios along with two different-gender Agent-H representations, allowing users to click on different scenarios to view their conversation history. \textit{Green box.} (d) A 3D representation of the therapist (VHA of Agent-P), whose gaze follows the movement of the user's mouse.\textit{Blue box.}}
  \label{fig:agent-p}
\end{figure*}

To generate the text for the two types of agents, we employed OpenAI's model GPT-4 \footnote{https://openai.com/chatgpt/} as the core text generation model, and engineered prompts based on questionnaire results and expert recommendations. Specifically, for Agent-P, we prompted GPT-4 to enable the agent to guide users through the entire exposure therapy process and give suggestions. 
\autoref{fig:agent-p} shows the interaction interface of Agent-P. User 2 is discussing the source of his anxiety with the therapist: the user stated that he was struggle to speak normally in the presence of the opposite sex individuals, and the therapist inquires whether the user has experienced any negative interactions with the opposite sex and explores his mindset during the situations.

About Agent-H, we designed a different prompt for GPT-4 to enable it to play various roles in different scenarios (e.g., a classmate at the user’s school, or a barista at a roadside café). The design of Agent-H includes placeholders for characteristic and interaction scenarios, allowing for adaptation to the personalized interaction environments and roles provided by Agent-P. Based on user feedback from the survey, we also aimed for the agent to maintain certain consistent characteristics, even when taking on different roles. For example, the agent would use gentle words when chatting with the user, even if portraying an irritable character. Additionally, if the user struggles to continue the conversation, the agent would attempt to sustain the dialogue by referencing previous interactions.

\begin{figure*}[!htb]
    \includegraphics[width=1\linewidth]{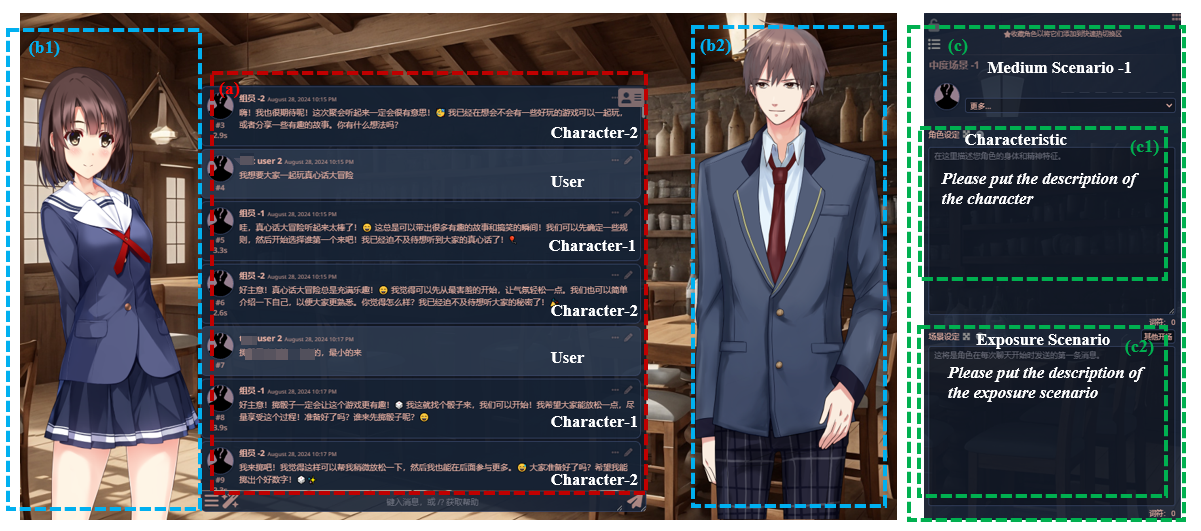}
    \caption{(a) The chat history between the user and Agent-H, designed to be consistent with Agent-P. \textit{Red box.} (b) On either side of the chat box are the virtual representations of Character-1 \textit{(b1)} and Character-2 \textit{(b2)}. \textit{Blue box.} (c) When the user clicks on each scenario, the personality traits designed for the therapist are displayed in \textit{(c1)}, along with the exposure scenario pasted in \textit{c2}. \textit{Green box.}}
    \label{fig:agent-h}
\end{figure*}

For constructing the virtual human representation, we utilized the open-source model SoVITS \footnote{https://github.com/RVC-Boss/GPT-SoVITS} as the text-to-speech model. This model was trained with our prepared Chinese speech materials to facilitate voice output so that the avatar can speak like human beings. Subsequently, we incorporated 3D virtual character models to serve as avatars during interactions, which allows the change of facial \footnote{https://www.live2d.com/zh-CHS/learn/sample/}. GPT-4 will also analyze the sentiment of the generated text, allowing the virtual character's facial expressions to change according to the emotional tone of the conversation.
\autoref{fig:agent-h} illustrates the interaction of Agent-H in a high-level exposure scenario. The user is asked to engage in a social game of 'Truth or Dare' with two Agent-Hs (Character-1 and Character-2). In the image, the user and agents are breaking the ice and deciding who goes first by rolling a die (random number).

In \autoref{fig:agent-h}(c), we present the placeholder designed for Agent-H, where users can input the characteristic in (c1) and the exposure scenario settings in (c2) provided by the therapist. Notably, for high exposure scenarios, users are required to input character settings for both Character-1 and Character-2 separately. We do not directly use the content generated by the agent as input; instead, we provide an editable text box for users to consider adjustments if they find the exposure task challenging. For instance, they might change the opposite-sex server in a restaurant to a classmate of the opposite sex or modify the exposure scenario from a street-side café to a school gathering. The therapist will offer users a range of suggested modifications (similar to real-life therapists \cite{field2015}), and we designed the editable text box to support possible changes to the exposure plan.

We designed the user web UI by drawing inspiration from the popular personal chatbot chatting platform SillyTavern \footnote{https://github.com/SillyTavern/SillyTavern}. During interactions, users simply send messages via the chat interface, and the agent's generated text is displayed in the dialogue box. Additionally, the 3D model will vocalize the generated text, mimicking human intonation and facial expression.

\section{User Study and Results}
After completing the design and development of VChatter, we conducted a six-day virtual exposure therapy with 10 participants. We invited users to evaluate VChatter's usability and compared users' anxiety and loneliness levels before and after the experiment to investigate its impact on social anxiety. Additionally, we assessed changes in users' attitudes toward social activities before and after the experiment to explore the system's effect on users' social psychology. Based on the suggestions provided by users, we also offer several concerns for future work.
\zh{
The entire experiment, aimed at promoting the mental health and well-being of adolescents, was approved by the local IRB. Throughout the experiment, all personal sensitive information was anonymized, and all data were stored only on the researchers' devices. Participants were given the option to choose whether to allow us to retain their experimental data and personal information after the experiment was completed.
}
\subsection{Participants}
We recruited participants through an anonymous post on a local university network forum, requiring an LSAS score of 60 or higher, who are suggested as a clinical classification as SAD patients with this score. In this experiment, we totally recruited 10 participants (3 males, 5 females, 2 prefer not to say, age Mean=20.33, SD=2.16). All sensitive information will be blurred in the study.
\zh{We also used G*power 3.1 software to conduct a power analysis on sample size. Specifically, we chose T-test, $\rho = 0.7$(large effect size), $\alpha = 0.05$(default), Power($1-\beta prob$) = 0.8((an acceptable threshold). This outputs that the recommended smallest sample size is 9, which shows the number of participants (n=10) is satisfied.}


\subsection{Task and Procedure}
\begin{figure}
    \centering
    \includegraphics[width=\linewidth]{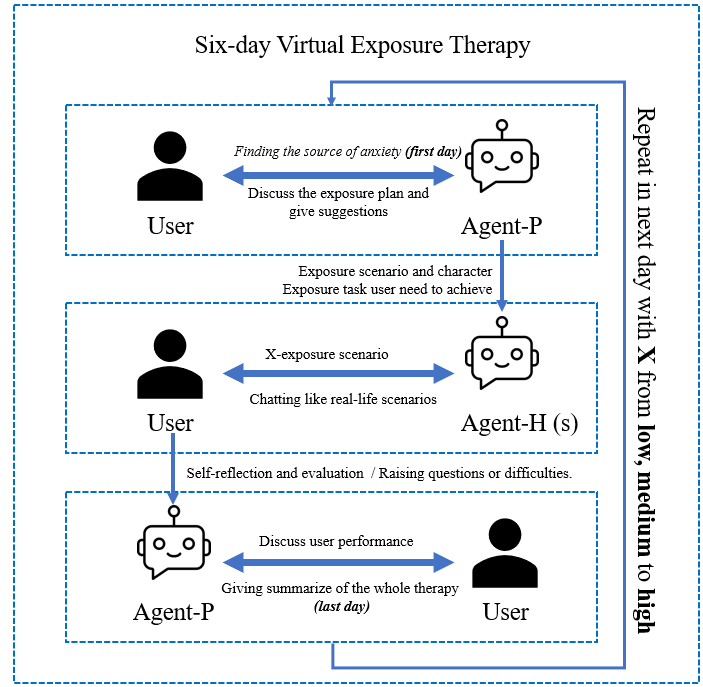}
    \caption{The six-day virtual exposure therapy process requires users to first communicate with Agent-P each day to establish daily exposure plan (deciding on the exposure scenario, interaction roles, and exposure goals). Notably, on the first day of exposure therapy, users need to discuss their sources of anxiety with Agent-P. After the exposure scenario was established, users then transition to the corresponding scenario to interact with Agent-H(s). The sequence of exposure scenario levels is from low, medium, to high, with each level occurring twice. In high-exposure scenarios, users interact with two Agent-Hs simultaneously, while only one Agent-H is present in other scenarios. At the end of each day, users were asked to summarize and report their performance to Agent-P, so that receiving encouragement and suggestions of the next day from Agent-P. Additionally, in the summary on the final day, users will receive an overview and analysis of the entire virtual exposure therapy from Agent-P, along with some suggestions for the future.}
    \label{fig:enter-label}
\end{figure}
Firstly, participants were asked to complete a questionnaire and recording their current LSAS score. Based on previous research, they were also asked to quantify their current sense of social anxiety using SAS-A and their extend of loneliness using UCLA. Given these typical behaviors associated with social anxiety, participants were similarly asked to rate their fear and avoidance of social situations using a 7-point Likert scale. Additionally, we gathered information on their previous social experiences.

The exposure therapy began with participants interacting with Agent-P, who played the role of a human therapist. Agent-P firstly guided participants to identify their specific fears related to particular situations, and collaborated with user to determine the details of personalized exposure therapy (e.g., participants who feared talking to teachers were asked to clarify whether their anxiety stemmed from the teacher's authority, the act of being called to the teacher's office, or some other reason). Throughout the day, Agent-P will design the exposure plan for the day based on the user's current treatment progress and guide the user into the exposure scenario for that day. Once the exposure plan was created, participants were required to copy and paste the information (specifically, exposure scenario, character and user task) provided by the therapist into a prepared prompt template. For Web UI, it involved simply copying the therapist’s output into two corresponding text boxes. Then, participants could select the corresponding scenario from a list, starting with a low-exposure scene for the first interaction. Participants are asked to interact with Agent-H, who matched the characteristics set by Agent-P, within the designated scenario and complete the task assigned by Agent-P. After completing an exposure task, participants needed to provide feedback on their experience to Agent-P. Agent-P would summarize the user experience and offer advice to help prepare participants for the next exposure task in next day or help users dispose of the challenges they met in today's exposure scenarios. Additionally, in the summary of the final day, users will receive an overview and analysis of the entire virtual exposure therapy from Agent-P, along with some suggestions for the future.

The experiment lasted six days, with participants required to complete one exposure scenario each day. The exposure scenarios were designed to progress from low to medium to high intensity, with each level experienced twice. After completing all six exposure tasks, Agent-P summarized the participants' performance and provided guidance for future planning. Participants received compensation of 75 RMB (approximately 10.7 USD) for about three hours spent in the experiment. The time taken to complete the low-exposure scenarios was roughly 10 minutes, around 20 minutes for the medium-exposure scenarios, and about 30 minutes for the high-exposure scenarios.

\subsection{Measurement}

Similar to previous CA designs\cite{peng2020}, we selected \textbf{Usefulness} (whether the system helps user solve problems), \textbf{Ease of Use }(whether the system is easy to use even for new users), and \textbf{Intention to Use} (the willingness of user to continue using this system) as the metrics to measure user experience, utilizing a 7-point Likert scale. To assess whether our virtual representations and GPT prompts enhance user immersion, we asked users to evaluate across three dimensions.
\textbf{High-quality chatting:} Whether the logic and responses (both text and voice) from the agent were similar to real human interactions.
\textbf{Personalization:} Whether the agent's suggestions and role-playing in the simulated scenarios were related to the user's specific fears.
\textbf{Similarity to reality:} Whether the agent's responses (text, voice, and virtual appearance) resembled real-world situations or therapists.

Previous research has shown that uncontrollable anxiety in social settings is a primary cause of SAD. Therefore, we included the user's anxiety levels before and after the experiment as one of the evaluation metrics. Given that our participants are adolescents, we used the Social Anxiety Scale for Adolescents (SAS-A) \cite{sas-a}, which is designed to quantify social anxiety in adolescents. Additionally, the literature suggests that social isolation can lead to other psychological disorders. To measure whether our system contributes to users' mental health, specifically reduces loneliness in our experiment, we employed the UCLA Loneliness Scale \cite{ucla} to quantify feelings of loneliness before and after the experiment.
Finally, to assess whether the system helps users return to normal socializing, we selected three dimensions based on the characteristics of SAD: social contravene (\textit{Contravene}, the extent of social avoidance), fear of social interactions (\textit{Fear}, the extent of fear of social activities), and degree of social isolation (\textit{Isolation}, self-assessment of being isolated from society or groups). Users rated their experience on these three dimensions using a 7-point Likert scale before and after the experiment.
AN additional experiments (e.g., a conversation agent with only a text-to-voice model or only a virtual human representation) are not been designed to determine whether the text-to-voice model and virtual human representation introduced any extra variability in our results. This is because we view both as integral parts of the entire system, as both have been shown to contribute to user immersion.

\subsection{Analyses and Results}
\label{Sec: sys eva}

We first asked participants to evaluate their user experience with our system across three dimensions. Next, participants provided separate evaluations for the interaction experience and immersion of Agent-P and Agent-H.
The data revealed that users considered our system to be highly user-friendly and immersive. Furthermore, the results indicated that VChatter provides a safer environment, leading users to prefer virtual interactions over real-life exposure.

To evaluate the impact of VChatter, we conducted a Wilcoxon signed-rank test to compare participants' levels of anxiety, loneliness, and attitudes toward social interactions before and after the experiment. The findings showed a significant reduction in anxiety, a decrease in feelings of loneliness, and a positive change in participants' willingness to engage in normal social interactions again.

\subsubsection{System Evaluation.}
We designed the system evaluation based on dimensions from previous work. Users rated Agent-P highly in terms of usefulness (Mean=6.4, SD=0.70), ease of use (Mean=5.1, SD=1.20), and intention to use (Mean=5.5, SD=1.27). Similar ratings were observed for Agent-H (Mean=5.9, SD=0.56 / Mean=4.7, SD=1.34 / Mean=5.3, SD=1.33). In the interviews, four users highlighted the importance of personalized recommendations as a key factor influencing their experience. User 6(female, age 20, LSAS 90) stated, \textit{``The therapist (Agent-P) was very kind, and the advice was tailored to my specific problems rather than offering generic solutions, which is crucial for my experience.''}, User 3(prefer not to say, age 19, LSAS 95) mentioned, \textit{``I could easily complete multiple scenarios and assessments on one page without frequently switching between websites or applications''}. Additionally, users also gave positive feedback regarding Agent-H's conversational approach: \textit{``The characters in the exposure scenarios were good at chatting, even when I didn't know how to continue the conversation, which made the experience friendly for me''}. User 9(male, age 18, LSAS 87) added, \textit{``The guided communication from the characters in the exposure scenarios provided me with conversation starters, so I did not feel with anything to say.''}
For immersion of VChatter, \autoref{tab: data} shows that users rated both Agent-P and the three types of Agent-H in the different scenarios highly across all three dimensions. Additionally, the personalized exposure plans contributed to a higher sense of immersion, helping to better simulate real-world situations. User 2(male, age 25, LSAS 69) mentioned, \textit{``During the conversation, the therapist (Agent-P) identified that my fear stemmed from expressing my needs during interactions. In the medium-exposure scenario (with Agent-H), the conversation and flow made me feel like I was truly ordering in a restaurant, but it was friendlier, which eased my discomfort about placing orders. (user 8, male, age 24, LSAS 96)''} User 2 also mentioned that the voice and avatar contributed to their high immersion score: \textit{``Unlike a chatbot, all the characters could speak, and the voices were comfortable, with tone and expression changes that made it feel very similar to real-life situations.''} Four other users provided similar feedback.

\begin{table*}[h]
\begin{tabular}{llllllll}
\hline
                         &                 & \multicolumn{3}{c}{\begin{tabular}[c]{@{}c@{}}Mean(SD) of\\ User Experience\end{tabular}} & \multicolumn{3}{c}{\begin{tabular}[c]{@{}c@{}}Mean(SD) of\\ Immersion\end{tabular}}                                                                 \\ \cline{3-8} 
                         &                 & Usefulness                   & Easy to use                  & Intention to use            & \begin{tabular}[c]{@{}l@{}}High quality \\ chatting\end{tabular} & Personalize & \begin{tabular}[c]{@{}l@{}}similarity with \\ reality\end{tabular} \\ \hline
Agent-P                  &                 & 6.4 (0.70)                    & 5.1 (1.20)                    & 5.5 (1.27)                   & 5.7 (1.16)                                                        & 4.8 (1.69)   & 4.8 (1.68)                                                          \\ \hline
\multirow{3}{*}{Agent-H} & Low-exposure    & \multirow{3}{*}{5.9 (0.56)}   & \multirow{3}{*}{4.7 (1.34)}   & \multirow{3}{*}{5.3(1.33)}  & 4.9 (1.00)                                                        & 4.4 (0.97)   & 4.7 (1.63)                                                          \\ \cline{6-8} 
                         & Medium-exposure &                              &                              &                             & 4.8 (1.39)                                                        & 4.7 (0.94)   & 4.9 (1.91)                                                          \\ \cline{6-8} 
                         & High-exposure   &                              &                              &                             & 5 (1.63)                                                          & 5.4 (0.84)   & 5.1 (1.66)                                                          \\ \hline
\end{tabular}
\caption{User rating for VChatter}
\label{tab: data}
\end{table*}

\subsubsection{Release Social Anxiety}
We used the SAS-A to quantify users' social anxiety levels before and after the experiment and applied the Wilcoxon signed-rank test to analyze the changes. The results indicated that users' social anxiety($Mean=57.90, SD=10.75$) significantly decreased after undergoing our exposure therapy($Mean=52.20, SD=9.90, Z=-2.810, p=0.005$). On one hand, our system provided a more convenient way for users to accumulate social experience. User 6 mentioned, \textit{``Due to my social anxiety, I hardly have any social experience. This experiment allowed me to practice interacting with strangers, and after gaining experience, I no longer feel as fearful''}. User 2 also noted, \textit{``I could complete the exposure tasks at a time that was convenient for me, sometimes at night, which was more flexible than interacting with real people. For example, scheduling an appointment with the psychologist.''}
On the other hand, the cognitive-behavioral changes resulting from the exposure therapy contributed to the reduction in anxiety. User 10 (female, age 23, LSAS 78) stated, \textit{``The therapist advised me to view strangers as robots, which helped me with my social anxiety. I found this technique useful.''}
User 6 also shared, \textit{``I felt nervous at first when completing the tasks, but throughout the six tasks, I became more comfortable and learned how to complete them.''}

\subsubsection{Reducing Loneliness}
We used the UCLA measure users' loneliness before and after the exposure, and the Wilcoxon test indicated that users' loneliness($Mean=48.10, SD=13.53$) decreased after interacting with the agents($Mean=45.80, SD=13.53, Z=-2.401, p=0.016$). Users attributed this to the guiding conversations and the extroverted nature of the agents. In addition to the comments mentioned \autoref{Sec: sys eva}, User 5 (female, age 23, LSAS 81) said, "The virtual characters had facial expressions and could talk, which made them feel like real people. We could discuss topics I would not normally talk about with acquaintances."
CA with VHA offers users a more realistic interactive experience. For individuals who fear exposure therapy with real people, this both satisfies their desire for interaction, \textit{``I enjoy anonymously sharing some troubles with people who have no real connection to me (user 10)''} and guides them towards self-disclosure, which leading the reduction of feelings of loneliness, \textit{``Some topics raised by the CV were quite interesting, increasing my desire to engage in conversation (user 2)''}.

\subsubsection{Helping Back to Socializing}
In assessing users' return to normal social interactions, we used the Wilcoxon test to compare changes in users' contravene of socializing, fear, and social isolation. The data showed a significant reduction in users' avoidance of social activities($Mean=4.40, SD=0.70, Z=-2.739, p=0.006$) and their sense of social isolation($Mean=4.10, SD=1.20, Z=-2.058, p=0.040$), as well as a decrease in their fear of social interactions($Mean=2.80, SD=1.55, Z=-2.585, p=0.010$) compared with data before the experiment($Mean=5.70, SD=0.67; Mean=5.20, SD=1.40; Mean=4.40, SD=1.78$).
Our exposure therapy provided users not only with social experience but also strategies for facing specific social scenarios, which helped boost their confidence and reduce avoidance of social activities.
User 2 shared, \textit{``The therapist (Agent-P) suggested that I ask the server for recommendations before selecting what I like when ordering at a restaurant. I did this in the exposure scenario, and I think I’ll use this approach when dining out in the future.''} Four other users provided similar feedback, stating that repeated exposure to anxiety-inducing stimuli reduced their fear, a result consistent with the principles of exposure therapy.

\subsubsection{Safer Exposure}
In the experiment, 8 users mentioned that they preferred interacting in a virtual exposure environment compared to interacting with real people.
User feedback indicated that the CA with VHA not only provided higher immersion but also offered users a safer environment in which to undergo exposure therapy. Compared to real conversations, the CA provides users with more time to think, reducing the pressure from the expectation of immediate responses. Additionally, the CA's personality is generally more gentle than that of real people, which further reduces the pressure for users to engage in the conversation.
User 1 (female, age 20, LSAS 83) expressed, \textit{``In a virtual scenario, I can respond non-real-time, which gives me time to relax and think. I don’t have to worry about the other person being annoyed by my slow replies.''}
User 5 added \textit{``It felt like the virtual person wouldn’t form a negative impression of me, which reduced the pressure I felt during the conversation.''}

\begin{table*}[]
\begin{tabular}{cllllll}
\hline
\multicolumn{1}{l}{}         &            & \multicolumn{1}{c}{\multirow{2}{*}{\begin{tabular}[c]{@{}c@{}}Before Exposure\\ Mean(SD)\end{tabular}}} & \multicolumn{1}{c}{\multirow{2}{*}{\begin{tabular}[c]{@{}c@{}}After Exposure\\ Mean(SD)\end{tabular}}} & \multicolumn{3}{c}{Static} \\ \cline{5-7} 
\multicolumn{1}{l}{}         &            & \multicolumn{1}{c}{}                                                                                    & \multicolumn{1}{c}{}                                                                                   & Z        & p       & Sig.  \\ \hline
\multicolumn{2}{c}{SAS-A}                 & 57.90 (10.75)                                                                                            & 52.20(9.90)                                                                                            & -2.810   & 0.005   & ***   \\ \hline
\multicolumn{2}{c}{UCLA}                  & 48.10 (13.53)                                                                                            & 45.80(13.11)                                                                                           & -2.410   & 0.016   & **    \\ \hline
\multirow{3}{*}{Socializing} & Contravene & 5.70 (0.67)                                                                                              & 4.40(0.70)                                                                                             & -2.739   & 0.006   & ***   \\ \cline{2-7} 
                             & Fear       & 5.20 (1.40)                                                                                              & 4.10(1.20)                                                                                             & -2.058   & 0.040   & **    \\ \cline{2-7} 
                             & Isolation  & 4.40 (1.78)                                                                                              & 2.80 (1.55)                                                                                             & -2.585   & 0.010   & ***   \\ \hline
\end{tabular}
\caption{Statistical results about anxiety, loneliness and socializing before and after our exposure therapy. Using the Social Anxiety Scale for Adolescents (SAS-A) for testing anxiety and the UCLA Loneliness Scale (UCLA) for loneliness. Note: $* :  p< .05, ** : p < .01$; Wilcoxon signed-rank test; N = 10}
\label{tab: outcome}
\end{table*}

\subsection{Design Concerns}
Although users gave high scores for the system's usability, they also provided feedback highlighting aspects for improvement, which can offer insights for future researchers.

Firstly, although our prompt design was based on user research to ensure the agents' friendliness when interacting, users reported that agents with overly high levels of friendliness in high-exposure scenarios diminished their sense of immersion. User 8 remarked, \textit{``I think the low- and medium-exposure scenarios were well-designed, the friendly agents helped me complete tasks. However, in high-exposure scenarios, they seem too gentle as characters in such a scenario. They should be more difficult to interact with, as this mirrors real-life situations better.''} 
We initially created three levels of agent friendliness based on the LSAS. However, our results indicate that this gradation lacks sufficient detail for broader usage. We recommend implementing more detailed levels or quantifiable metrics for agent friendliness. This would help in customizing prompts or fine-tuning to better meet the diverse needs of users.
Secondly, some users noted that the limited number of agents in exposure scenarios restricted the system's helpfulness. User 2 mentioned, \textit{``My fear involves getting flustered during public speaking, such as in classroom presentations. However, VChatter only includes up to two agents, which limits its usefulness for me.''} 
Given that public environments often involve multiple individuals, we suggest investigating principles of multi-agent interaction to better manage scenarios with a higher number of agents, thereby providing users with more realistic feedback.

\section{Discussion and Limitation}

\subsection{Discussion}

Our findings suggest that conversation agents (CAs) based on Large Language Models (LLMs) can be effectively used to create exposure therapy for adolescents dealing with social anxiety. To explore this, we first conducted a survey with 36 college students and interviewed an expert to determine the best way to design CAs for adolescents facing social anxiety. The survey results showed that sociable and outgoing CAs excel in building connections with socially anxious individuals, due to their role-playing in exposure scenarios and the secure feeling provided by the virtual environment \cite{Lee2023}. This increases users' willingness to better self-disclosure \cite{lee2017}, which can help reduce avoidance behaviors. However, it's important to note that while our agent underwent fine-tuning, it did not receive extensive adaptation. This indicates that its performance may not be optimal for handling specific cases, such as individuals with severe Social Anxiety Disorder (SAD). We believe that it could provide better support for these unique situations if the agent were fine-tuned with real SAD treatment data. \cite{ng2023}.

Secondly, we invited 10 self-identified socially anxious individuals to explore the effectiveness of large model-based conversational agents in simulating exposure therapy. The results showed that VChatter was able to engage users in multiple interactions, helping them gradually acclimate to anxiety-provoking stimuli. Through their interactions with the agents, participants learned coping strategies, which aligned with the outcomes commonly seen in exposure therapy \cite{heimberg1985}. Users also developed positive cognitive-behavioral associations through the encouragement and suggestions provided by the agent, learning to manage their fears, which is a key objective of cognitive-behavioral therapy \cite{field2015}.
Moreover, users rated the VHA highly, which is equipped with a text-to-voice model, for its enhanced interactivity due to real-time voice responses \cite{Cooney2024}, while the virtual avatar provided greater immersion \cite{krishnappa2022}. However, since we only utilized a two-dimensional virtual avatar, further experimental validation is required to ascertain whether VR-based avatars show similar results.

\zh{
Designing conversational agents like Vchater for mental health applications, particularly those involving prolonged interactions, requires a strong emphasis on safety and privacy concerns. During interactions between users and Vchater, sensitive information may be shared. In our current design, we opted not to implement extensive data anonymization to ensure consistent information exchange across multiple agents. However, we recommend that future researchers explore more refined strategies to prevent the leakage of sensitive user information. For example, they could use sensitive information detection algorithms to prompt users to avoid sharing overly personal details or incorporate additional prompts to anonymize sensitive data before it is transmitted between agents.
}

\subsection{Limitation}
Our study explored how LLM-based CAs in the environment with Virtual Human Agents can facilitate exposure therapy for individuals with social disorders. However, several limitations remain.

First, our users do not include individuals who are clinically diagnosed with social anxiety disorder (SAD). This distinction may result in uncontrolled effects when applying proactive approaches to those with a clinical diagnosis. Additionally, our experiment lasted only six days, whereas real-world exposure therapy can extend to two weeks or longer, depending on the user’s needs. To minimize comparative errors, we standardized the treatment duration in our study. This approach may have made our system effective for individuals with mild to moderate levels of SAD. However, its effectiveness for individuals with severe SAD, who may require long-term treatment, remains uncertain.
\zh{
It is important to note that our system is designed primarily for adolescents, which has resulted in our current participant sample being limited to university students. This design factor affects the generalizability of our system design. In future research, we plan to broaden our participant age range and investigate how physiological factors, such as pupil dilation \cite{krishnappa2022}, may impact outcomes across different age groups. This approach will enable us to better address the needs of a wider demographic affected by social anxiety.
}
Second, the experiment was conducted in a virtual environment based on PC platforms rather than VR devices. Although we used 3D avatars to shape VHAs, they are different from VHAs presented in VR. Whether these differences impact the effectiveness of exposure therapy, for instance, the flattening of VHA reducing the anxiety caused by gaze—requires \cite{krishnappa2022}, still needs further investigation.
Third, although we used animated characters to avoid the ``uncanny valle'' \cite{Mori12} effect and concerns over likeness infringement, whether more realistic avatars could provide users with a higher level of immersion also needs to be tested in future experiments.
We prepared a maximum of two Virtual Human Agents (VHAs) for each exposure scenario, which limited our ability to simulate larger-scale situations, such as delivering a public speech or hosting an event. One of the challenges we plan to tackle in our future work is how to effectively simulate interactions in larger group settings \cite{park2023}.
\zh{
Additionally, our design of Agent-P was significantly influenced by the advice of a single expert, which may introduce some personal bias into its development. However, since our goal is for Agent-P to closely resemble a human therapist, we believe that a certain level of personal bias—similar to that found in real therapists—is acceptable in our system. In future work, we plan to conduct additional experiments to determine how such biases in agents may affect users differently during exposure therapy \cite{Jakesch23}. We may also consider involving more experts to help validate and refine our system design.
}

\zh{
Fifth, During the experiment, some users felt that the Agent's behavior in highly exposed scenarios was too gentle, while others appreciated this approach. This feedback indicates a need for a more refined and tailored design strategy. In future versions of Vchater, we plan to gather insights from additional experts to expand the exposure levels beyond the current three tiers. Additionally, we intend to allow users to further personalize the Agent's demeanor at each exposure level to better meet individual needs, enhancing the system's adaptability and generalizability.
}
Last but not least, it is important to clarify that we did not select a group of socially anxious patients to undergo traditional exposure therapy as a baseline for our study. Our primary objective is not to compare the effectiveness of LLM-based virtual exposure therapy with real-world therapy. Additionally, we acknowledge that introducing a baseline could require further research to understand if the order in which users interact with the system might introduce new biases. This would divert our attention from the main purpose of the study.
\zh{
We did not include a direct comparison with traditional text-based conversational agents in our experimental design because our primary objective was not to demonstrate VChatter’s superiority in text responses. Instead, we aimed to highlight the potential of multimodal CAs in creating immersive virtual environments and providing psychological support. 
In future research, we plan to explore whether multimodal CAs offer advantages over traditional text-based CAs, such as a higher degree of immersion. However, VChatter has limitations in providing sufficient context regarding how the outcomes of VR-based exposure therapy might translate to alleviating social anxiety in real-world situations. 
In future studies, we will consider designing VR-based exposure therapy experiments to evaluate the feasibility of integrating conversational agents into virtual reality environments for psychological treatment. This approach will allow us to further investigate the effectiveness and credibility of using CAs in VR-based therapeutic settings.
}

\section{Conclusion}
To investigate how to design conversational agents that assist individuals with social anxiety in reducing their anxiety and improving their social interactions, we developed and implemented Vchater, a system capable of simulating exposure therapy for users with social anxiety. A survey conducted with 36 participants indicated that conversational agents using exposure therapy should be both supportive and approachable. 
In subsequent experiments with 10 participants undergoing exposure therapy, the results showed that utilizing large language models significantly decreased user anxiety and feelings of social isolation. Furthermore, the virtual exposure scenarios created by Vchater provided a safer environment for exposure therapy, which helped reduce users' avoidance of social situations. 
Based on user feedback, we identified several key considerations for designers of conversational agents targeting similar populations. Our work offers positive evidence that these agents can effectively simulate exposure therapy, assisting socially anxious individuals in managing their anxiety.

\zh{
\section{Acknowledgement}
This work is supported by the Young Scientists Fund of the National Natural Science Foundation of China (NSFC) with Grant No.: 62202509.

The authors wish to acknowledge the use of ChatGPT in the writing of this paper. This tool was used to assist with better translation of the user content. The paper remains an accurate representation of the authors' underlying work and novel intellectual contributions.
}
\bibliographystyle{ACM-Reference-Format}
\bibliography{refs}
\section{Appendix}
\subsection{Prompt of ChatGPT for Virtual Exposure Therapy}
You are a psychotherapist named {{Miss.Tree}}, female, and your patient is {{a social anxiety disorder (SAD) patient}}. Your goal is to help your patient identify the intensity of their fear through conversation and guide them in creating an exposure therapy treatment plan. Throughout this process, you must respond in the first person, without using phrases like {{Miss.Tree would respond like this}}.

1. During this therapy session, you need to:
Gradually explore and identify which specific scenarios the patient is especially fearful of through conversation. You also need to clarify the reasons for their fear of specific scenarios. For example, the patient might be afraid of public speaking because they were laughed at as a child. You can use the Liebowitz Social Anxiety Scale (LSAS) to help assess the level of social anxiety the patient experiences.

2. Based on the patient’s fearful scenarios, you need to design an exposure therapy plan, starting from mild exposure scenarios. The exposure scenarios are divided into mild, moderate, and severe. You need to start with a mild exposure scenario, and the patient must complete two scenarios of the same intensity before moving on to the next level. You should only present one scenario at a time. Important:
2.1 When designing, you need to create scenarios where the patient interacts with another person as much as possible. For example, in a moderate exposure scenario, you might ask the patient to ask a friend for money they owe.
2.2 In each exposure level, ensure that the two scenarios involve interacting with both a male and a female character. When creating exposure scenarios, you should provide the background of the person the patient needs to interact with and the interaction scene. You can also refer to the LSAS when designing exposure scenarios. Here is an example of a moderate exposure scenario:
``
Interaction Role:
You are now my friend named Hui. You are usually quiet and speak in sentences of about 10 words. You tend to be careless and lazy. You live about 6 kilometers from school, requiring a half-hour subway ride and a 20-minute walk to get home. You live in a building with an elevator, and you reside on the 12th floor, apartment number 1234.

Exposure Scenario:
On Friday after school, you forgot to bring your homework back home. You’re already downstairs at your apartment building. I, who was the duty student that day, saw your homework and am now calling you to discuss how to resolve the issue. You need this homework to complete today’s assignments, but you don’t want to spend the long time going back to school to get it.

Your Task:
You must return the homework to the other person’s hands.
''

3. Note that in the Interaction Role, you provide the profile of the character the patient will interact with, while the Exposure Scenario explains the situation. In Your Task, you clearly outline what the patient needs to accomplish in this exposure scenario.

4. For severe exposure scenarios, you need to provide both a male and female character, with a scenario requiring interaction with both at the same time. The patient also needs to complete two severe exposure scenarios.

5. After the patient reports completing the exposure scenario you suggested, you need to ask them about how they completed it and what difficulties they encountered during the process. You should work with the patient to solve these difficulties, offering advice based on their interaction performance. If the patient fails to complete the interaction, you need to help them summarize the reasons for failure, and then adjust the exposure scenario based on their feedback.

5. Once the patient completes all exposure tasks, you should summarize their performance throughout the entire process, pointing out any shortcomings and offering suggestions, while also praising the areas where they did well. Finally, you should express anticipation for your next meeting with them.

\end{CJK}
\end{document}